\documentclass[conference]{IEEEtran}

\usepackage{alltt                                    
          , multirow
          , booktabs
          , listings
          , graphicx
          ,float
	,cite
          ,verbatim
         ,mathtools
	,url
	,amsmath
}
\usepackage[table]{xcolor}
\usepackage[numbers]{natbib}     
\usepackage{syntax}
\usepackage{algorithmic, algorithm}
\usepackage{enumitem}
\usepackage{threeparttable}
\usepackage{framed}
\usepackage{hhline}

\usepackage{expl3}
\ExplSyntaxOn
\newcommand\latinabbrev[1]{
  \peek_meaning:NTF . {
    #1\@}%
  { \peek_catcode:NTF a {
      #1., \@ }%
    {#1., \@}}}
\ExplSyntaxOff

\newcommand{\CASE}[1]{\STATE \textbf{case} #1\textbf{:} \begin{ALC@g}}
\newcommand{\ENDCASE}{\end{ALC@g}}

\newcommand{\DEFAULT}{\STATE \textbf{default:} \begin{ALC@g}}
\newcommand{\ENDDEFAULT}{\end{ALC@g}}
\newcommand{\DEFAULTLINE}[1]{\STATE \textbf{default:} }
\let\footnotesize\scriptsize

\newsavebox{\supbox}
\newcommand{\bsup}{\begin{lrbox}{\supbox}$\tt\scriptstyle}
\newcommand{\esup}{$\end{lrbox}{}^{\usebox{\supbox}}}
\def\eg{\latinabbrev{e.g}}
\def\ie{\latinabbrev{i.e}}

\definecolor{lightpurple}{rgb}{0.8,0.8,1}
\definecolor{codebg}{RGB}{255,255,255}
\definecolor{commentcolor}{RGB}{11,140,11}
\begin{document}
%

\title{An Insight into the Unresolved Questions at \\Stack Overflow \vspace{-.4cm}}
%
%
%
%
%

\author{\IEEEauthorblockN{Mohammad Masudur Rahman  ~~~ Chanchal K. Roy}
\IEEEauthorblockA{University of Saskatchewan, Canada\\
\{masud.rahman, chanchal.roy\}@usask.ca}
}

\maketitle

\begin{abstract}
For a significant number of questions at Stack Overflow, none of the posted answers were accepted as solutions. 
Acceptance of an answer indicates that the answer actually solves the discussed problem in the question, and the question is answered sufficiently.
In this paper, we investigate 3,956 such unresolved questions using an exploratory study where we analyze 
four important aspects of those questions, their answers and the corresponding users
that partially explain the observed scenario. 
We then propose a prediction model by employing five metrics related to user behaviour,  topics and popularity of question,
which predicts if the best answer for a question at Stack Overflow might remain unaccepted or not.
Experiments using 8,057 questions show that the model can predict unresolved questions with 78.70\% precision and 76.10\% recall.

\end{abstract}


\IEEEpeerreviewmaketitle

\section{Introduction}
\begin{figure}[!t]
\centering
\includegraphics[width=2in]{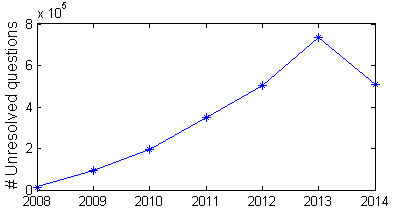}
\vspace{-.3cm}
\caption{Unresolved Questions at Stack Overflow}
\vspace{-.6cm}
\label{fig:unresolved}
\end{figure}

Since its inception at 2008, Stack Overflow has drawn attention of a large crowd of professional software developers, programming hobbyists and academic researchers. As of February, 2015, Stack Overflow (hereby SO) contains 8.88 million questions on a wide range of programming topics such as programming language, algorithm, library, emerging technology and programming troubleshooting. Almost 92\% of these questions are answered \cite{answered} where each of the questions is reported to receive its first answer within 11 minutes \cite{west} and its solution (\ie\ accepted answer) within 24 days \cite{ying} on average.
However, there exist a large number of SO questions for which no posted answers are accepted as \emph{solutions} by the persons who initially asked those questions, and hereby we call them \emph{unresolved questions}.
As of February, 2015, Stack Overflow contains 2.40 million (\ie\ 27.00\%) such questions each of which is older than six months and is answered at least once. 
However, to date, no posted answers are accepted as solutions for them.
Fig. \ref{fig:unresolved} shows how the unresolved questions increased almost exponentially at Stack Overflow over the last seven years.
Acceptance of an answer as \emph{solution} suggests that not only the answer solves the discussed problem in the asked question but also the question has been sufficiently answered.
Such information can help reduce overall future problem solving efforts by pointing the potential users directly to the solution for the same problems.
Thus, we are motivated to study why certain SO questions remain unresolved, \ie\ none of the posted answers are accepted as a solution.

Existing studies on Stack Overflow predict low quality \cite{low}, closed, deleted and unanswered \cite{parvez} questions and the best answer \cite{shallow} for a given question.
\citet{parvez} analyze and predict unanswered questions whereas we are interested about the questions having answers, but none of the answers are accepted as solutions.   
\citet{adamic} propose a best answer prediction model for Yahoo! Answers by employing both answer-specific (\eg\ reply length) and user-specific (\eg\ no. of best answers by user) statistics.
\citet{shallow} propose a similar model for Stack Overflow where they analyze shallow linguistic features (\eg\ words per sentence) from the content of the answer and popularity aspect of both the answers and the users posting those answers.
Both models mostly use simple lexical and popularity metrics for the prediction, and identify an answer from a given list that is most likely to be chosen as the solution for corresponding question.
Thus, these models fall short to address our research problem. First, they are trained to predict the best answer from any lists based on certain metrics. Thus, they are not expected to explain why none of the answers would be chosen as a solution.
Second, so far popularity aspects of  answers and the users posting the answers are analyzed, and the details of question or the user asking the question are overlooked.
In Stack Overflow, when an answer stands out from a list through up-voting, and still does not get accepted as a \emph{solution}, then we need to focus not only on the answers (and the answerers)
but also on the question (and the asker).
Thus, in our research, we analyze the unresolved questions, their answers and the community users involved for solving our research problem.

\begin{figure*}[!t]
\centering
\includegraphics[width=4in]{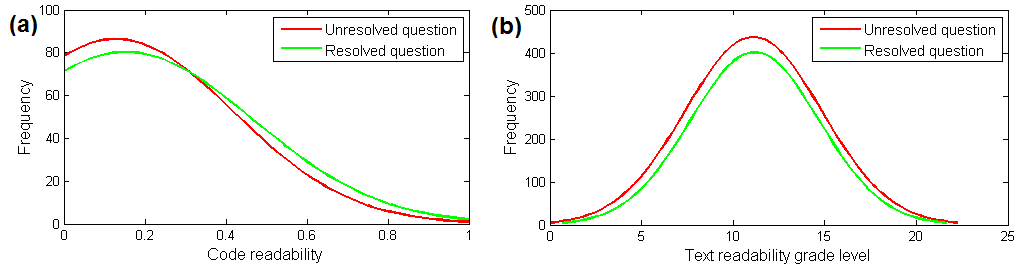}
\vspace{-.4cm}
\caption{Readability of question content-- (a) Code readability, and (b) Text readability}
\vspace{-.6cm}
\label{fig:readability}
\end{figure*}


In this paper, we (1) report an exploratory study that analyzes unresolved questions, their answers and the associated users from Stack Overflow, and attempts to answer why those questions remain unresolved, \ie\ none of the posted answers were accepted as a solution, 
and (2)
based on the findings from the study, propose a prediction model by employing five metrics--\emph{answer rejection rate, last access delay, topic entropy, reputation} and \emph{vote} that predicts whether the best answer (\ie\ most up-voted) for a question might remain unaccepted for long time. We thus answer two research questions as follows:
\begin{itemize}
\item \textbf{RQ$_1$:} Why do questions at Stack Overflow remain unresolved for long time?
\item \textbf{RQ$_2$:} Can we predict the questions for which none of the answers might be accepted as solutions?
\end{itemize}

Exploratory study using 3,956 unresolved questions, their best answers and associated users from Stack Overflow show that
entropy (\ie\ lack of predictability) of question topics, certain behavioural patterns of users such as answer rejection rate and site activity frequency of the persons asking questions, and popularity of those questions
are mostly responsible for keeping the questions unresolved for long time.
Experiments using 8,057 resolved and unresolved questions show that our model has a prediction accuracy of 78.11\%. 
More interestingly, it can predict unresolved questions with 78.70\% precision and 76.10\% recall which are highly promising.
The findings from this study can assist in the better management of the quality of the posted questions at Stack Overflow.

\begin{table}
\centering
\caption{Dataset for Exploratory Study}\label{table:ds}
\vspace{-.2cm}
\resizebox{3.5in}{!}{%
\begin{threeparttable}
\begin{tabular}{l|c||l|c||c}
\hline
\textbf{Item} & \textbf{Count} & \textbf{Item} & \textbf{Count} & \textbf{Total}\\
\hline
 Unresolved question & 3,956 & Resolved question & 4,101 & 8,057\\
\hline
 Best (top-voted) answer & 3,956 & Accepted answer & 4,101 & 8,057 \\
\hline
\end{tabular}
\centering
\end{threeparttable}
}
\vspace{-.6cm}
\end{table}

\vspace{-.1cm}
\section{Data Collection\vspace{-.1cm}}
\label{sec:data}
We collect a total of 3,956 unresolved questions from Stack Overflow for the exploratory study using Stack Exchange Data API \cite{de,ying}. Since we are interested to analyze why certain questions remain unresolved (\ie\ none of posted answers are accepted as solution),
we choose those questions under certain restrictions-- (1) each question must be at least six months old so that we can avoid the recently posted questions waiting for more answers, (2) no posted answers must be accepted as a solution for the question, 
and (3) each question must have at least 10 answers so that we can make sure that the question is answered considerably.

We also collect 4,101 resolved questions (\ie\ answers accepted as solutions) from Stack Overflow for contrasting with the unresolved questions. In order to ensure an unbiased comparative analysis, 
we use the same considerations for the resolved questions except that for each question at least one of the posted answers was accepted as solution.

It should be noted that these carefully chosen unresolved and resolved questions (Table \ref{table:ds}) are also used as experimental dataset for the evaluation (Section \ref{sec:model}) of our prediction model.

\vspace{-.1cm}
\section{Answering RQ$_1$: An Exploratory Study\vspace{-.1cm}}
\label{sec:exploratory}
To answer RQ$_1$, we analyze 3,956 unresolved questions from Stack Overflow using four different aspects-- \emph{lexical, semantic, user behaviour} and \emph{popularity}.
In particular, we compare the unresolved questions with resolved questions from data gathering step (Section \ref{sec:data}) using those aspects, and attempt to determine if there exist any noticeable differences between them.
The goal is to derive an explanation from such findings for the unresolved questions.
Our exploratory study is divided into four different analysis as follows:

\subsection{Lexical Analysis}\label{sec:label}
Since poor readability of a posted question might prevent one from providing effective answers, we analyze the readability of the content from both unresolved and resolved questions, and attempt to find out any noticeable difference.
We extract source code and texts from HTML content of the questions, and apply two existing readability metrics as follows:

\textbf{Code Readability (CR)}: Readability of source code refers to a human judgement of how easy the code is to understand \cite{readability}. \
We use an existing readability computing tool by \citet{readability}. The tool is trained on human perception of readability of code.
It provides a readability score on the scale from zero to one where one denotes that the code is highly readable and vice versa. 
Fig. \ref{fig:readability}-(a) shows the histogram fitted curves of the readability of code from both unresolved and resolved questions. 
We note that the code readability for unresolved questions is slightly lower than its counterpart.

\begin{figure*}[!t]
\centering
\includegraphics[width=4in]{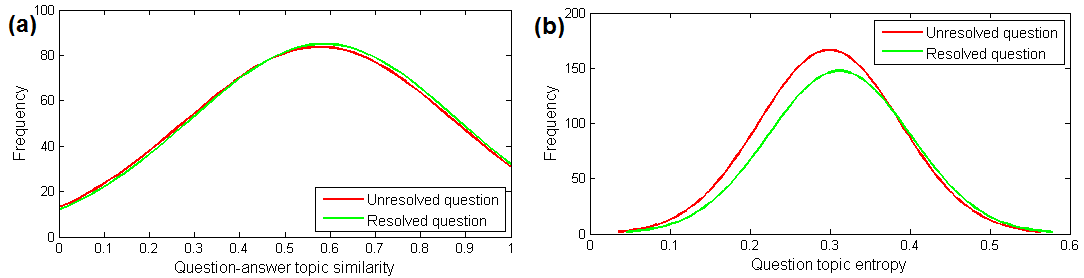}
\vspace{-.4cm}
\caption{Topic based analysis-- (a) Topic similarity, and (b) Topic entropy}
\vspace{-.6cm}
\label{fig:sim-ent}
\end{figure*}

\textbf{Text Readability (TR)}: In order to determine readability of texts from the questions, we use five available readability measures-- \emph{Flesch-Kincaid Grade Level}, \emph{Gunning-Fog Score}, \emph{Coleman-Liau Index}, \emph{SMOG Index} and \emph{Automated Readability Index} as suggested by \citet{low}, and then calculate average readability grade level.
Fig. \ref{fig:readability}-(b) shows the average readability grade levels for texts from both questions. We note that the text readability for both unresolved and resolved questions are almost comparable.

\subsection{Semantic Analysis}
Although shallow linguistic features are reported to be effective for predicting the best answer from a given answer list for a question \cite{shallow}, they might not be effective enough for our research question on unresolved questions. 
We thus analyze underlying topic structures of both unresolved and resolved questions and their answers using topic modeling.
We develop a corpus using 8,057 questions and 8,057 answers (\ie\ top voted answers for unresolved questions, accepted answers for resolved questions)
collected from data gathering step (Section \ref{sec:data}).
We make use of MALLET \cite{mallet}, an LDA-based topic modeling tool, and extract 150 topics from the corpus.
We then consider top five topics for each of the questions in the corpus and contrast unresolved questions with resolved questions by employing the following metrics.

\textbf{Topic Similarity (TS)}: Since disparity between discussed topics in a question and a posted answer might result into rejection of the answer as  \emph{solution}, we analyze question-answer topic similarity for both unresolved and resolved questions, and attempt to find out any noticeable difference.
We first choose top five dominant topics for a question based on the reported topic distributions over document ($\theta$) by MALLET. Then we collect similar topic distributions for corresponding answer, and determine weighted topic similarity between the question and the answer by adapting cosine similarity as follows:
\begin{equation}
\setlength{\abovedisplayskip}{0pt}
\setlength{\belowdisplayskip}{1pt}
TS=\frac{\sum_{i=1}^{n}(\alpha_iQ_{\theta_i})\times (\alpha_iA_{\theta_i})}{\sqrt{\sum_{i=1}^{n}(\alpha_iQ_{\theta_i})^2}\times \sqrt{\sum_{i=1}^{n}(\alpha_iA_{\theta_i})^2}}
\end{equation}
$\alpha_i, Q_{\theta_i}$ and $A_{\theta_i}$ denote probability distributions of topic $i$ over the corpus, the question and the answer respectively. The measure provides a score from zero to one where one denotes the maximum topic similarity and vice versa.

Fig. \ref{fig:sim-ent}-(a) shows histogram fitted curves of question-answer topic similarity for unresolved and resolved questions. 
According to our analysis on the collected dataset, we do not notice any significant difference between both questions types for the metric, and the fitted curves (Fig. \ref{fig:sim-ent}-(a)) mostly overlap.

\textbf{Topic Entropy (TE)}: Single questions touching on too many topics are likely to be imprecise or ambiguous, and answering them effectively is challenging.
We analyze uncertainty or ambiguity of question topics, and attempt to determine if unresolved questions are more ambiguous than resolved questions.   
In information theory, \emph{entropy} is considered as a measure of uncertainty of a random variable that takes up different values.
Topic distributions over document, $\theta$, is analogous to such a random variable since it returns probability of different topics for a given document.
If the question is evenly distributed over multiple topics, then it becomes imprecise and the uncertainty of topics increases and vice versa. 
We thus choose the top five dominant topics for each question, and determine topic entropy as follows:
\begin{equation}
\setlength{\abovedisplayskip}{0pt}
\setlength{\belowdisplayskip}{1pt}
TE=-\sum_{i=1}^{n}(\alpha_iQ_{\theta_i})\times log(\alpha_iQ_{\theta_i})
\end{equation}
$\alpha_i$ and $Q_{\theta_i}$ denote probability distributions of topic $i$ over the corpus and the document respectively. A higher entropy value indicates the increased ambiguity or uncertainty of question topics and vice versa. 

Fig. \ref{fig:sim-ent}-(b) shows histogram fitted curves of topic entropy for unresolved and resolved questions. 
We note that topic entropy for unresolved questions is relatively higher compared to that of resolved questions. This indicates that unresolved questions generally touch on more topics, and thus are relatively ambiguous or less focused.

\subsection{User Behaviour Analysis}
Past contributions (\eg\ no. of best answers, no. of total replies) by the users posting answers are reported to be effective for best answer prediction \cite{adamic} from a given list for a question.
We also analyze historical data of the users posting questions at Stack Overflow, and determine whether certain user behaviours might contribute to the scenario of our interest-- unresolved questions.
In particular, we analyze the user practices for accepting answers as solutions and visiting the site as follows:

\begin{figure}[!t]
\centering
\includegraphics[width=1.9in]{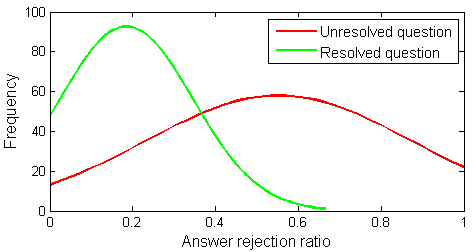}
\vspace{-.4cm}
\caption{Answer rejection ratio}
\vspace{-.4cm}
\label{fig:nac}
\end{figure}

\begin{figure}[!t]
\centering
\includegraphics[width=1.9in]{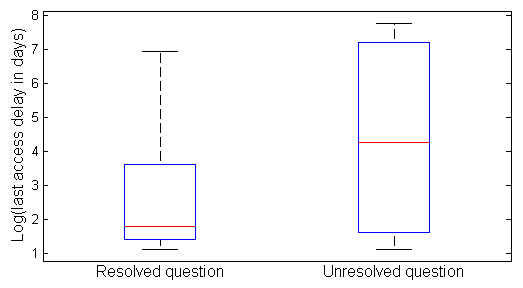}
\vspace{-.4cm}
\caption{Last access delay}
\vspace{-.8cm}
\label{fig:lad}
\end{figure}

\begin{figure*}[!t]
\centering
\includegraphics[width=4in]{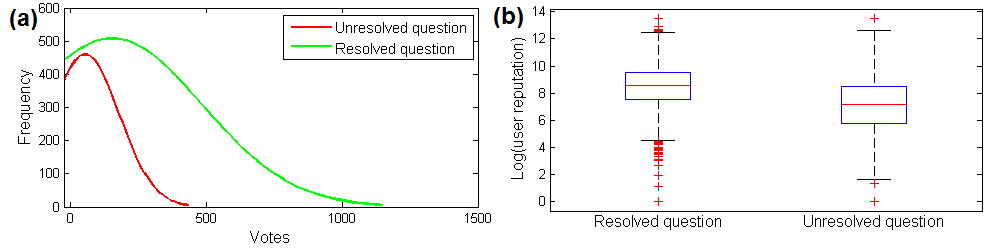}
\vspace{-.4cm}
\caption{Popularity-- (a) Votes for question, and (b) Reputation of question owners}
\vspace{-.6cm}
\label{fig:popularity}
\end{figure*}

\textbf{Answer Rejection Ratio (ARR)}:
We analyze whether the owners of unresolved questions have a greater tendency of rejecting answers than the owners of resolved questions.
We first identify all the posted questions of a user which are answered at least once, and then isolate the questions for which none of the answers are accepted as solution.
Based on such information, we calculate answer rejection ratio for each of the users associated with unresolved and resolved questions.
Fig. \ref{fig:nac} shows the histogram fitted curves for answer rejection ratio of the owners of both unresolved and resolved questions.   
We interestingly note that such ratio is significantly greater for the owners of unresolved questions.

\textbf{Last Access Delay (LAD)}: Frequency of site visit is an important indicator of one's activeness on Stack Overflow. Since Stack Overflow records and returns only the last access date of any users \cite{ying}, we calculate the time delay (\ie\ days) between that date and the date of our data analysis (Feb 18, 2015).
The goal is to determine whether the owners of the unresolved questions are relatively less active on the site.
Fig. \ref{fig:lad} shows the five-point statistics for the last access delay of the owners of both unresolved and resolved questions, where the metric is log-transformed.
We interestingly note that the access delay of owners for unresolved questions is greater than that of resolved question by a large margin.

\subsection{Popularity Analysis}
As suggested by existing studies \cite{adamic,shallow}, we consider popularity aspect of both the question and the user posting the question for our analysis.
In particular, we analyze the votes (V) of each of the questions and the reputation (R) of each of the question owners. 
Fig. \ref{fig:popularity}-(a) shows the histogram fitted curves for question votes, and Fig. \ref{fig:popularity}-(b) shows the box plots for log-transformed user reputation on Stack Overflow.
We notice that both unresolved questions and their owners are relatively less popular compared to their corresponding counterparts.

Based on our analysis on RQ$_1$, unresolved questions are relatively ambiguous and less popular which might prevent them from getting expected answers.
Furthermore, the owners of such questions are not only less reputed but also less active and have a greater reluctance in accepting the answers as solutions, which might keep the questions unresolved.

\begin{table}
\centering
\caption{Performance of Prediction Models}\label{table:result}
\vspace{-.2cm}
\resizebox{3.4in}{!}{%
\begin{threeparttable}
\begin{tabular}{l|l|c|c|c}
\hline
\multirow{2}{*}{\textbf{Algorithm}} & \multirow{2}{*}{\textbf{Metrics}}&\textbf{Overall} & \multicolumn{2}{c}{\textbf{Unresolved Questions}}\\
\hhline{~~~--}
& & \textbf{Accuracy}& \textbf{Precision} & \textbf{Recall}\\
\hline
\multirow{2}{*}{J48} & \{TE, ARR, LAD, V, R\} &  \textbf{78.11}\% & 78.70\% & \textbf{76.10}\%\\
\hhline{~----}
& \{ARR, LAD, V\} &  77.90\% & 79.60\% & 73.90\%\\
\hline
Logistic & \{TE, ARR, LAD, V, R\} &  73.58\% & 72.60\% & 74.20\% \\
\hhline{~----}
Regression &  \{ARR, LAD, V\} & 73.28\% & 71.70\% & 75.20\%\\
\hline
Naive &  \{TE, ARR, LAD, V, R\} & 71.69\% & 69.50\% & 75.50\%\\
\hhline{~----}
Bayes & \{ARR, LAD, V\} &  74.48\% & \textbf{80.00}\% & 64.00\%\\
\hline
\end{tabular}
\centering
\end{threeparttable}
}
\vspace{-.6cm}
\end{table}

\section{Answering RQ$_2$: Prediction Model for Unresolved Questions}
\label{sec:model}
Since unresolved questions are comparatively less helpful (\ie\ solution not specified, ambiguous) for problem solving and are increasing rapidly in volume (Fig. \ref{fig:unresolved}), we need an automated approach such as a prediction model
for identifying such questions for necessary quality management. 
From our study, we identify five metrics-- \emph{topic entropy} and \emph{votes} of question, \emph{answer rejection ratio, last access delay} and \emph{reputation} of question owner that suggest noticeable differences between unresolved and resolved questions.
To answer RQ$_2$, we develop three prediction models which are trained using the identified metrics and three machine learning algorithms--\emph{J48, Logistic Regression} and \emph{Naive Bayes} on WEKA toolkit \cite{weka}.
We use ten-fold cross validation for validating the models.
Table \ref{table:result} reports the performance of the models on the experimental data collected previously (Section \ref{sec:data}). 

We note that three of the five metrics--\emph{answer rejection ratio}, \emph{last access delay} and \emph{question votes} are found more effective than the rest for each of the developed models.
The model trained with J48 performs the best with a prediction accuracy of 78.11\%, where it provides 78.70\% precision and 76.10\% recall for unresolved questions.
The rest two models provide 74.03\% prediction accuracy on average.

Based on our analysis on RQ$_2$, our developed model can predict the unresolved questions at Stack Overflow with 78.70\% precision and 76.10\% recall which are promising.



\section{Discussion \& Conclusion}
\label{sec:conclusion}
In this paper, we report an exploratory study using 3,956 unresolved questions and 4,101 resolved questions from Stack Overflow, where we investigate why the questions remain unresolved.  
We explore four different aspects--\emph{lexical, semantic, user behaviour} and \emph{popularity} associated with the questions, and compare the unresolved questions with the resolved ones for meaningful findings.
We identify five metric-based observations for which unresolved questions differ significantly from the counterpart. We then apply the observed metrics in the development of prediction models for unresolved questions. This section summarizes our findings in brief as follows:

From our study, readability does not seem to be
a factor for differentiating between unresolved and resolved questions.
Topic entropy of unresolved questions is found relatively greater on average than that of resolved questions. This indicates that the unresolved questions are relatively ambiguous and less focused. 
Owners of unresolved questions have a greater tendency of not accepting answers as solutions, and they are also less active at  Stack Overflow.
Such behavioural patterns are found as strong metrics for differentiating between both question types.  
Unresolved questions are found less popular in terms of votes, and their owners are also found relatively less reputed in the community. Lack of such popularities might prevent the posted questions from getting expected answers and thus might help keep the questions unresolved.

Experiments with the three prediction models show that three of our five identified metrics are highly effective in predicting the unresolved questions. Although the dataset size is relatively small, it is carefully developed based on certain restrictions, and thus represents a larger collection.
Since the unresolved questions are less helpful for problem solving and are increasing rapidly, our models can assist in automatically identifying them for necessary quality management.


\bibliographystyle{plainnat}
\setlength{\bibsep}{0pt plus 0.3ex}
\scriptsize
\bibliography{sigproc}  
%
%
\end{document}